# Analysis and Enhancements of Leader Elections algorithms in Mobile Ad Hoc Networks

Mohammad H. Al Shayeji[1], AbdulRahman R. Al-Azmi[2], AbdulAziz R. Al-Azmi[3] and M.D. Samrajesh[4]
Computer Engineering Department, Kuwait University, Kuwait
[1]alshayej@eng.kuniv.edu.kw, [2]raphthorne@yahoo.com, [3]fortinbras222@hotmail.com, [4]sam@differentmedia-kw.com

*Abstract*— Mobile Ad Hoc networks (MANET), distinct from traditional distributed systems, are dynamic and self-organizing networks. MANET requires a leader to coordinate and organize tasks. The challenge is to have the right election algorithm that chooses the right leader based on various factors in MANET. In this paper, we analyze four leader election algorithms used in mobile Ad Hoc Networks. Factors considered in our analysis are time complexity, message complexity, assumptions considered, fault tolerance and timing model. Our proposed enhancements include recovered nodes inquiring about the current leader and the use of candidates during election to reduce the overhead of starting a new election session. In addition, better election criteria specific to MANET, such as battery life and signal strength, are proposed. Our evaluation and discussion shows that the proposed enhancements are effective. The analysis can be used as a reference for system designers in choosing the right election algorithm for MANET.

## I. Introduction

Distributed systems are the backbone of modern day computing services. An election algorithm elects a leader to coordinate and organize tasks in distributed systems that includes Mobile Ad Hoc networks (MANET). In contrast to traditional distributed systems, nodes arrive and leave MANET more frequently. In the case of a leader node departure or failure, nodes detecting the non-availability of the leader initiate a leader election process to select a new leader. This election process should be completed in a finite number of steps with a consensus among the nodes on the new leader [24].

Leader election is a primary control problem in both wireless and wired systems [1]. In wireless networks, leader has a variety of applications including key distribution [2], routing coordination [3], sensor coordination [4], and general control [5][6].

In this paper, we analyze four leader election algorithms applied in MANET using common factors such as time complexity, message complexity, assumptions considered, fault tolerance and timing model. Our proposed enhancements include waking up nodes that send an inquiry message to their functioning neighbors to identify the current leader instead of initiating a costly new election. In addition, we propose utilizing candidates during election to reduce the overhead, and use better election criteria specific to MANET such as battery life and mobility. Moreover, we propose to use the CSMA/CD algorithm to handle multiple diffusion computations since it requires fewer number of messages.

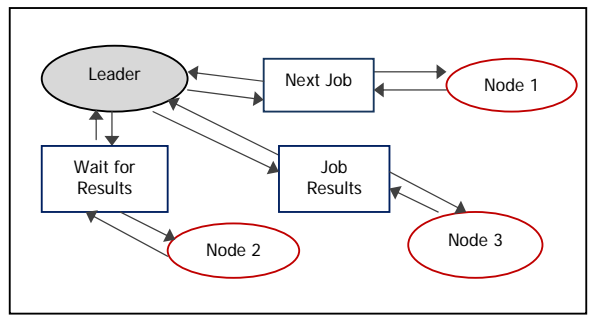

Figure 1. A Typical leader coordination

The organization of the paper is as follows: section II describes the related work, and section III provides background information on leader election algorithms. The factors for our analysis are described in section IV, section V presents the analysis and the proposed enhancements, and evaluation and discussion on the analysis is presented in section VI. Finally, in section VII, we present our concluding remarks and suggestions for future work.

## II. Related Work

Despite leader election being a classical problem, few researchers have given importance to the leader election process in context of MANET. Leader election algorithms have different assumptions as some algorithms use synchronous message transfer[8], and others use asynchronous message transfer[7][8]. The leader election process has been studied and compared based on its complexities in [9], however the algorithms are limited to distributed systems.

A survey on Service Discovery Protocols (SDP) in MANET is presented in [10]. The survey discusses the features of several SDPs ; however, the study is not specific to leader election algorithms.

MANET ability to continue its task uninterrupted in spite of attacks or intrusions is presented in [11]. The paper identified the survivability properties of MANET and recommended that a survivable MANET needs to consider a multi-layer and multi-attack solution. However, the paper deals only with the security aspect of MANET.

A survey on use of Genetic Algorithms (GA) in MANET is discussed in [12]. The paper does a comparative study between two GA-based algorithms, however the survey concentrated only on the QoS routing in Ad-Hoc Networks.

Most of the above work focuses on the general aspects of MANET while our work focuses on the specific leader election problem in MANET. Additionally, our analysis and enhancement provide a reference point for system designers to choose a right election algorithm in MANET.

### III. BACKGROUND

#### A. General Leader election

Distributed systems require a coordinator/leader to coordinate and organize tasks such as implementing mutual exclusion. Leader election is required when the system is initialized, or the leader crashes. A leader election algorithm elects a leader from among the nodes considering application-specific conditions. The key to the leader election problem is attributed to G. LeLann [13]. He formalized a method to create a new token when the token is lost in a ring network. The token holder can transmit on the medium, and this token holder represents leader that coordinates tasks in distributed systems. Primary leader election algorithms are the Bully Algorithm and Invitation Algorithm [8].

#### B. Mobile Ad Hoc Network

MANET is an Ad Hoc network, Ad hoc is a Latin phrase meaning "for this" [14] and no topology exists among the communicating nodes that are connected [15]. MANET is highly decentralized and need a coordinator to organize their work. The coordinators are leaders of the sub-network that control routing, distributing the workload, and synchronizing the other nodes. Applications of MANET include wireless sensor networks, military and industrial unmanned robots, and more.

### IV. FACTORS FOR ANALYSIS

We establish common factors for the analysis of the four algorithms. These factors reflect major design aspects that specify the advantages for each algorithm [16][17].

#### A. Complexities

Time and message complexities depend on the number of participating nodes.
- Time Complexity: The number of steps required to reach consensus on the newly elected leader measures time complexity.
- Message Complexity: The number of messages required by nodes to elect a new leader measures message complexity.

#### B. Assumptions

Assumptions include all the pre-conditions required for the algorithm to work. Some algorithms assume reliable channels [18], FIFO (First in First Out) channels, or nodes that know information about their neighbors.

#### C. Fault Tolerance

Discussions on fault tolerance techniques are presented in [19][20][6]. Most MANET algorithms must deal with network partitioning and merging.

#### D. Timing Models

Timing in distributed systems is not global as there is no global clock; distributed algorithms use synchronous assumptions to achieve synchrony [18][3][21].

### IV. ANALYSIS AND ENHANCEMENT

#### A. Research of Asynchronous Leader Election Algorithm on Hierarchy Ad Hoc Network[22]

*1) Analysis*

Initially the network establishes a hierarchy within election groups. When a node is separated from the network during the election, it is not allowed to join the election process. However, if the node sends a *'hello'* message, receiving nodes compare it with its Cluster ID and will accept or deny. The paper uses extrema finding diffusion computation and leaders are chosen considering the system's most beneficial choice. Moreover, this approach does not consider the case, in which leader node is separated during the election process. The response is to have another election.

*2) Enhancements*

a) Nodes create clusters based on their locality and closeness in vicinity. Elections are conducted in tiny clusters, and then clusters' newly elected leaders broadcast their identity to other clusters similar to the approach in Garcia's Invitation Algorithm [8]. The performance is better since the initial leader election is in small groups.

b) Un-visited nodes should not be allowed to join the election process until it is completed since this increases the overhead on election.

#### B. Design and Analysis of a Leader Election Algorithm for Mobile Ad Hoc Networks[1]

*1) Analysis*

The algorithm is based on diffusion computation and extrema finding, and it handles topological changes dynamically. Several nodes can start the algorithm but a node can only participate in one diffusion computation. Diffusion computation will search for a node with highest value, which is determined by the node's Unique Identity (UID). The algorithm works in an asynchronous environment with channels that are not FIFO.

When any node detects the loss of its leader, the algorithm starts. The initiator sends an '*election'* message that spreads to all its neighbors thus spanning out as a tree growing. The other nodes, upon the arrival of the '*election'* message, forward it to their children and declare the sender as their parent. When the source receives an acknowledgement, it diffuses a '*leader'* message to the entire newly constructed spanning tree with the information of the current new leader.

*2) Enhancement*

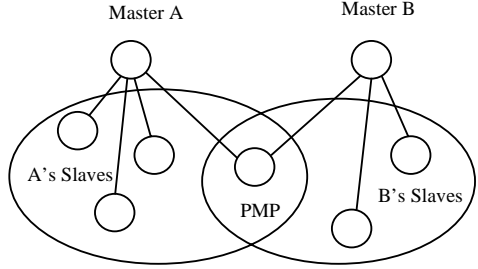

Figure 2. Participant in Multiple Piconets (PMP)

a) Algorithm assumes when a node wakes up from crashing, it should start an election to identify its new leader. This is costly. In the proposed enhancement, waking nodes send an inquiry message to their functioning neighbors and identitfy the current leader instead of a starting a new election.

b) The algorithm uses extrema finding using the nodes UIDs in which a node aborts its current diffusion and joins the higher order UID. A better method is to resolve the comparison based on the diffusion starter's quality as this resolves the process more rapidly.

### C. A *Leader Election Algortihm within Candidates on Ad Hoc Mobile Networks[23]*

*1) Analysis*

The paper is an extension of [8] and [20]. The assumptions include nodes with UIDs, asynchronous FIFO bidirectional channels, and sufficiently large node buffers, The algorithm also deals with networks partitions and mergers that can occur during the election process. The addition to the original algorithms [8][20] is that it uses candidate lists. Candidate lists contain the UIDs of current leader and four other main candidates.

When nodes check that the leader's heartbeats have stopped, the nodes initiate election messages. Multiple messages are handled by checking the message with the highest UID attached to it. When this election message reaches the leaf nodes in the diffusion computation-spanning tree, they send acknowledgement messages to their parents. These acknowledgment messages contain the candidate lists. On the way back to the initiator node, the lists are updated. Update is done to obtain the highest known UID of the nodes. Finally, the initiator combines the candidate lists and the highest UID is elected as the new leader.

*2) Enhancement*

a) We recommend that election of leaders must use the highest remaining battery life and the lowest mobility range in the election process. Since nodes with low mobility have least chances of merging and partitioning from their current clusters.

b) The algorithm uses the UID to handle the multiple diffusion computations to elect a new leader. We recommend using CSMA/CD algorithm to handle such multiple diffusion computations since it requires *less number of messages.*

### D. Cooperative Leader Election Algorithm for Master/Slave Mobile Ad Hoc Network[26]

*1) Analysis*

An election algorithm for master/slave networks is proposed in [26]. The master is responsible for assigning and coordinating tasks and roles including scheduling mutual exclusive access. The slaves follow the master's orders; the algorithm assumptions are that by utilizing the algorithm iteratively, the MANET will end up with one unique leader. The system is synchronous, and its channels are reliable. Nodes have unique IDs.

In the election algorithm, each node maintains a tuple of two values, the ID of the current leader, and the current leader's master value. Nodes that wish to be the new master, replace their tuple with their own values, ID and criterion value. The master exchanges its tuple with each of its slaves. If it finds a node with higher criterion value, it will replace its own and broadcast it to the entire network of its slaves. In the network, there can be nodes that are Participant in Multiple Piconets (PMP) - a piconet is a network with at least two nodes and, at most, one master. Fig.2 shows a PMP layout. These nodes participate in multiple networks, but during each time slot, they obey one master.

*2) Enhancements*

a) In the algorithm, during election only the master can communicate with the rest of the nodes. In the proposed enhancement, we recommend slave nodes be allowed to communicate during election. This reduces the election time.

b) If a node wishes to start an election process, it sends a message to its master or other active masters if in a PMP. If no master is currently active for the node, it broadcasts the need to its neighbors to start the election process, and nodes wishing to become leaders will make themselves candidates. This creates more chances for nodes to become leaders.

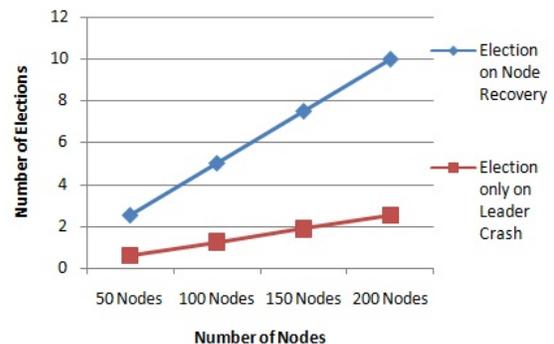

Figure 3. Number of Elections

TABLE I. Analysis Summary

| No | Paper | Complexity | | Assumptions | Fault Tolerance | Timing Models | Enhancements | Justification for Enhancements |
|---|---|---|---|---|---|---|---|---|
| | | Time | Message | | | | | |
| 1 | G. Zang [22] | O(n x (average delay)) | O(n) | Hierarchical networks, FIFO channels, UID | CSMA/CD for multiple elections, Handles partitions | AS | Nodes create cluster based on locality | The performance is better since the initial leader election is in small groups. |
| | | | | | | | Separated nodes should not join until end of election | New nodes introduce more overhead |
| 2 | Vasudevan [1] | $O(\log_{ANoN} PN+1^*)$ | $O(PN^{**})$ | Extrema finding, UID, No FIFO channels | Handles partitions, merges, and multiple elections | AS | Waking up nodes send an inquiry messages to find the leader | Re-election is costly and avoided |
| | | | | | | | Resolve based on the diffusion starter's | Rapid resolution |
| 3 | S. Lee, [23] | $O(\log_{ANoN} PN+1^*)$ | $O(PN^{**})$ | Extrema finding UID, No FIFO channels | Handles partitions, merges, and multiple elections. Has candidate list | AS | Leader selection based on better traits (e.g. battery life, mobility) | Nodes with low mobility have much less chances of merging and partitioning |
| | | | | | | | Use CSMA/CD | Fewer number of messages |
| 4 | R. Ali, [26] | 2 x ($\sum_{i \in P(G)} d(i)$ - P) + 1*** | 2 x ($\sum_{i \in P(G)} d(i)$ - P) + 1*** | Reliable channels, Role aware nodes | Nodes can leave and join the networks | SY | Participation of Slave nodes | Reduction in election time |
| | | | | | | | Send invitation messages | More chances for nodes to become leaders |

*: where *ANoN* is the *Average Number of Neighbors* in the participating nodes in the elections. *PN* is the *Participating Nodes* number.
**: *PN* is the *Participating Nodes* number during the elections. SY-Synchronous AS -Asynchronous
***: *Where P is the number of PMP nodes in the network G, P(G) is the set of PMP nodes in network G, and d(i) is degree of the PMP node i*.

## V. EVALUVATION AND DISCUSSIONS

The summary of our analysis is presented in Table-1. All the algorithms consider fault tolerance. Most MANET use asynchronous timing model.

### A. Comparative Evaluation

*1)* In [1], when crashed nodes wake up, they by default start an election. Accordingly, the number of elections conducted is directly proportional to the number of nodes waking up from crashes. In the proposed enhancement, nodes send inquiry messages to their neighbors to inquire about the current leader. This alternative would avoid full-scale elections. For example, if we consider 5% of nodes failed, we further assume that 25% of the failed nodes contain the leader. As shown in Fig 3, the number of elections conducted is considerably less and the cost of election is minimum.

*2)* In [23], when CSMA/CD is used before sending any message, the nodes wait for a certain timeout. When the ID is higher, the lower the timeout. Fig. 4 shows an scenario where node-1 fails and the message transfers. The dotted lines represent the case of using CSMA/CD and the number of messages are few.

*3)* Comparisons on [1] and [22] show [1] uses diffusion computation. This is better since the system can handle floods of messages. Moreover, when the system has more mobility, that leads nodes to leave the regions of connection and the locality of the leader. It would be better to choose [1] over [22] since the former allows multiple diffusions that may compensate the nodes' departures.

*4)* Comparisons on [25] and [26] show that both use extrema finding, yet the latter uses a valued node property in order to evaluate candidate nodes for leadership. Also, in the latter, the master can only exchange messages with the slave nodes while the former allows nodes to flood messages forming trees rooted at the initiator node. Finally, a major difference in [26], is that it is a synchronous algorithm while the rest are *asynchronous*

### B. Evaluation based on Factors

*1) Complexities*

The time and message complexities of all the algorithms presented in Table-I show they are linear. All algorithms are efficient; they only differ in the number of participating nodes or delay.

*2) Fault Tolerance*

Most algorithms handle partitions and merges. The use of FIFO channels is also present. In [23], candidates are used to avoid consecutive elections. We recommend using candidates in algorithms, since this reduce the overhead of starting a new election session.

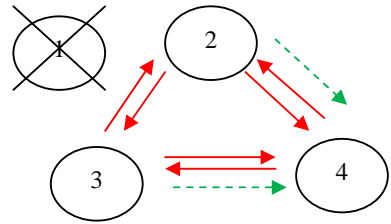

Figure 4. CSMA/CD in Leader election

*3) Assumptions*

Assumptions limit the algorithm from its wide applications hence the minimum assumptions lead to the greater the adaption of the algorithm for a real application.

*4) Timing Models*

Most MANET use asynchronous timing model. Nodes have no global time coordination among them.

## VI. CONCLUSION AND FUTURE WORK

We presented an analysis on leader election algorithm in MANET. The analysis results showed all algorithms analyzed considered fault tolerance in-depth. Our proposed enhancements include on recovery, nodes inquire about the current leader instead of initiating a new election process. In addition, we propose utilizing candidates during election to reduce the overhead, and use better election criteria specific to MANET such as battery life and mobility, using an asynchronous timing model is a better choice as most MANET are asynchronous. Moreover, we propose to use the CSMA/CD algorithm to handle multiple diffusion computations since it requires fewer number of messages

As a part of future work, we intend to include more recent leader election algorithms, additional factors for our further analysis, and empirical studies on proposed enhancements in the paper.

## REFERENCES

[1] S. Vasudevan, J. Kurose, and D. Towsley, "Design and Analysis of a Leader Election Algorithm for Mobile Ad Hoc Networks," ICNP'04, 2004.

[2] V. Park, and S. Corson, "A Highly Adaptive DistributedRouting Algorithm for Mobile Wireless Networks,"Proceedings to INFOCOM 97, pp. 1405–1413, 1997.

[3] N. Mohammed, H. Otrok, W. Lingyu, M. Debbabi, and P. Bhattacharya, "Mechanism Design-Based Secure Leader Election Model for Intrusion Detection in MANET," IEEE Transactions on Dependable and Secure Computing, vol.8, no.1, pp.89-103, 2011

[4] Allen Clement, Reaserch Statement, available: www.cs.utexas.edu/users/aclement/Allen_Research.pdf

[5] K. Hatzis, G. Pentaris, P. Spirakis, V. Tampakas and R. Tan. Fundamental Control Algorithms in Mobile Networks. In Proc. of 11th ACM SPAA, pages 251-260, March 1999.

[6] G. Singh, "Leader Election in the Presence of Link Failures," IEEE Transactions on Parallel and Distributed Systems, vol. 7, no. 3, March 1996.

[7] Amin Ansari, "Verification of Peterson's Algorithm for Leader Election in a Unidirectional Asynchronous Ring Using NuSMV," Cornell University Library, August, 2008, available: arXiv:0808.0962v1

[8] G. H. Molina, "Elections in a Distributed Computing System." IEEE Trans. Comp, vol.31, no. 1, pp.48-59, 1982.

[9] H. Abu-Amara, and A. Kanevsky, "On the Complexities of Leader Election Algorithms," Proceedings Fifth International Conference onComputing and Information ICCI, Page(s): 202, 29 May,1993

[10] A. Mian, R. Baldoni, and R. Beraldi, "A Survey of Service Discovery Protocols in Multihop Mobile Ad Hoc Networks," IEEE Pervasive Computing, 8 (1): 66-74, January-March 2009.

[11] M. Lima, A. dos Santos, and G. Pujolle, "A Survey of Survivability in Mobile Ad Hoc Networks," IEEE Communications Surveys & Tutorials, 11 (1): 66-75, First Quarter 2009.

[12] B. Kannhavong, H. Nakayama, Y. Nemoto, and N. Kato, "Application of Genetic Algorithms for QoS Routing in Mobile Ad Hoc Networks: A Survey," in the Proceeding of 2010 International Conference on Broadband, Wireless Computing, Communication and Applications BWCCA, pp. 250-256, November 2010.

[13] G. Le Lann. "Distributed systems: Towards a formal approach", Information Processing 77, Proc. of the IFIP Congress, pp. 155-160, 1977.

[14] Humayun Bakht, "History of Mobile Ad Hoc Networks", SCMS, Liverpool John Moores University, available: ww.oocities.org/humayunbakht/HMANET.pdf

[15] Advanced Network Technologies Division, "Wireless Ad Hoc Networks" National Institiue of Standards and Technology NIST, USA, available: www.antd.nist.gov

[16] Ali Ghodsi, "Distributed Algorithms 2g1513," available: www.sics.se/~ali/teaching/dalg/l06.ppt

[17] Mitsoa Valia, "Distributed Leader Election Algorithms in Synchronous Neworks," available: www.web.cs.gc.cuny.edu/~vmitsou/presentation.pdf

[18] Sukumar Ghosh, Distributed Systems: An Algorithmic Approach, Chapman & Hall/CRC, Tylor & Francis Group, LLC ,2007.

[19] M. Castro and B. Liskov, "Practical Byzantine Fault Tolerance and Proactive Recovery, ACM Transactions on Computer Systems," v. 20 n. 4, pp. 398-461, 2002.

[20] Leslie Lamport, "Time, Clocks, and the Ordering of Events in a Distributed System," Communications of ACM, vol. 21, no. 7, July 1978.

[21] S. Stoller, "Leader Election in Asynchronous Distributed Systems", IEEE Transactions on Computers, vol. 49, no. 3, March 2000

[22] Gang Zhang, Jing Chen, Yu Zhang, and Chungui Liu, "Research of Asynchronous Leader Election Algorithm on Hierarchy Ad Hoc Network," WiCom '09. 2009.

[23] S. Lee, M. Rahman, and C. Kim, "A Leader Election Algorithm Within Candidates on Ad Hoc Mobile Networks," Embedded Software and Systems, Lecture Notes in Computer Science, Vol. 4523, pp: 728-738, Springer Berlin / Heidelberg 2007.

[24] M. Pease, R. Shostak, and L.Lamport, "Reaching Agreement in the Presence of Faults," Journal of the Association for Computing Machinery, Vol 27, No 2, April 1980, pp 228-234

[25] N. Malpani, J. L. Welch, and N. Vaidya, "Leader Election Algorithms for Mobile Ad Hoc Networks," In the Proceedings of DIALM '00, NY, USA, 2000, ACM

[26] R. Ali, S. Lor, R. Benouaer, M. Rio, "Cooperative Leader Election Algorithm for Master/Slave Mobile Ad Hoc Network," 2nd IFIP Wireless Days (WD), Paris, 15-17 December 2009, pp. 1-5.